\documentstyle[aps,prb]{revtex}

\begin{document}
\draft
\title{ Scalar potential effect in an integrable Kondo model}
\author{Yupeng Wang$^{a,b}$ \and  Ulrich Eckern$^a$}
\address{$^a$Institut f\"{u}r Physik, Universit\"at Augsburg, 86135 Augsburg,
Germany\\
$^b$Laboratory of Ultra-Low Temperature Physics, Chinese Academy of Sciences, P. O. Box 2711, Beijing 100080, People's Republic 
of China}
\maketitle
\begin{abstract}
To study the impurity potential effect to the Kondo problem in a Luttinger liquid, we propose an integrable
model of two interacting half-chains coupled with a single magnetic impurity ferromagnetically.  It is shown 
that the scalar potential 
effectively reconciles the spin dynamics at low temperatures. Generally, there is a competition between the Kondo
coupling $J$ and the impurity potential $V$. When the ferromagnetic Kondo coupling dominates over the impurity potential
($V<|SJ|$), the Furusaki-Nagaosa many-body singlet can be perfectly realized. However, when the impurity potential
dominates over the Kondo coupling ($V\geq |SJ|$), the fixed point predicted by Furusaki and Nagaosa is unstable
and the system must flow to a weak coupling fixed point. It is also found that the effective moment of the impurity
measured from the susceptibility is considerably enlarged by the impurity potential. 
In addition, some quantum phase transitions 
driven by the impurity potential are found and the anomaly residual entropy is discussed.
\end{abstract}
\section{introduction}
With the development of the nanofabrication techniques for quantum wires and the prediction of edge states in the quantum 
Hall effect, the interest in one-dimensional (1D) quantum systems has been renewed in recent years\cite{1,2}. In fact, much
of the interest in 1D quantum systems is due to Anderson's observation\cite{3} that the normal state properties of the
quasi 2D high-$T_c$ superconductors are strikingly different from all known metals and can not be reconciled with Landau's
Fermi liquid theory but are more similar to properties of 1D metals. In another hand, the impurity problem has been a current 
interest in the field of condensed matter physics. A well known example is the Kondo problem, which stimulated a strong
challenge to traditional perturbation theory and provided a possible ``laboratory" to search for the non-Fermi liquid behavior.
The local perturbation problem to a 1D Fermi system has been the subject of an intensive theoretical investigation in the
recent years, for both its interesting anomalies with respect to that of a higher-dimensional system, and its relevance to
a variety physical situations such as the transport behavior of the quantum wires\cite{2,4,5} and the tunneling through a
constriction in the quantum Hall regime\cite{6}.
The Kondo problem in a Luttinger liquid 
 was first considered by Lee and Toner\cite{7}. Employing the perturbative renormalization
group method they found the power-law dependence of the Kondo temperature $T_k$ on the exchange coupling constant
$J$ with strong enough correlation among the conduction electrons. Subsequently, a
``poor man's" scaling treatment on this problem was performed by Furusaki and Nagaosa\cite{8}.
An interesting conjecture was proposed in their paper, which states that even
ferromagnetic Kondo coupling may induce Kondo screening at low temperatures. This
strong coupling fixed point describes a many-body singlet formed by the impurity and the
conduction electrons. 
They gave a simple picture to describe such a many-body
singlet. 
At first, one conduction electron will be bounded on the impurity site and forms 
a spin-1 composite with the impurity moment (spin-1/2). However, the other electrons on the
two nearest sites aside the impurity will screen this composite due to the antiferromagnetic exchange
induced by the correlation between the localized electron and the electrons near 
the impurity and therefore leave a spin singlet in the ground state. Moreover, they showed that the excess specific heat
and Pauli susceptibility due to the Kondo impurity show anomalous power law on the temperature, a
typical non-Fermi liquid behavior. The boundary conformal field theory has allowed
the classification of all possibilities of critical behavior for a Luttinger
liquid coupled to a Kondo impurity\cite{9} (without impurity potential). It turns there are only two
possibilities, a local Fermi liquid with standard low-temperature thermodynamics 
or a non-Fermi liquid with the anomalous scaling observed by Furusaki and Nagaosa.
We remark that there are still disputes\cite{10,11} on this problem and the situation
is not very clear yet.
\par

The impurity potential effect for the Kondo problem in 3D metals has been well understood. A weak impurity potential
 only renormalizes the
Kondo coupling $J$ by a factor $\cos^2\delta_V$ ($\delta_V$ is the phase shift of the scattering) but does not affect
the fixed point of the system\cite{12,13}. However, the magnetic impurity embedded in a 1D correlated electron
system with scalar impurity potential in addition to the Kondo coupling has been so far rarely studied\cite{14,15}.
As pointed out by Kane and Fisher\cite{5}, the impurity potential has a significant
effect in a Luttinger liquid. With the repulsive interaction among the conduction electrons,
it will be scaled to infinity at low energy scales and thus drives the system into two
disconnected half-chains. This observation has also been supported by the numerical studies of the finite size spectrum\cite{16}.
Such an effect is obviously detrimental to the formation of
Furusaki-Nagaosa singlet, since the impurity potential prevents the conduction electrons to occupy 
the impurity site. In fact, there must be a competition between the Kondo coupling $J$ and the scalar potential
$V$. When $J$ is dominant over $V$, the Furusaki-Nagaosa's strong coupling fixed point could be stable, 
otherwise the system may flow to other fixed points. Since the impurity cuts the whole chain into two pieces,
the Kondo problem in a Luttinger liquid is effectively a two-channel one\cite{17,18}.
Even without the scalar potential, the Furusaki-Nagaosa composite also prevents other electrons
to occupy the impurity site and thus behaves as an open boundary.
\par
Attempting to show how the impurity potential behaves for a magnetic impurity coupled ferromagnetically with a one-dimensional
host metal and how the Furusaki-Nagaosa conjecture realizes in a concrete system, we introduce an integrable model of 
two half-chains coupled with a magnetic impurity. We remark a few integrable impurity models have been introduced 
earlier
which describe either a transparent impurity in a whole chain\cite{19} or a half chain interacting with a boundary impurity\cite{14}. 
These models are disadvantageous to show the two 
channel Kondo physics and  can not capture all essentials of the Kondo problem in a Luttinger liquid. A good starting point 
is the following Hamiltonian, which describes a magnetic impurity in a Hubbard chain
\begin{eqnarray}
H=-t\sum_{j=-N, \sigma}^N [c_{j,\sigma}^\dagger c_{j+1,\sigma}+h.c]+U\sum_{j=-N}^N n_{j,\uparrow}n_{j,\downarrow}
+J\sigma_0\cdot{\bf S}+V(n_{0,\uparrow}+n_{0,\downarrow}),
\end{eqnarray}
where $J$ and $V$ are the Kondo coupling constant and the impurity potential, respectively; $\sigma_j$ is the spin operator
of the conduction electrons on site $j$ and ${\bf S}$ is the spin operator of the impurity. However, the above 
Hamiltonian is not exactly solvable and therefore can not provide us satisfactory information. Instead of to study (1), we
consider the following related Hamiltonian
\begin{eqnarray}
H=\sum_{\sigma=\uparrow,\downarrow}\int_0^L \frac{\partial c_{+,\sigma}^\dagger(x)}{\partial x}\frac{\partial c_{+,\sigma}(x)}{\partial x}dx+
\sum_{\sigma=\uparrow,\downarrow}\int_{-L}^0\frac{\partial c_{-,\sigma}^\dagger(x)}{\partial x}\frac{\partial c_{-,\sigma}(x)}{\partial x}dx+\nonumber\\
c\sum_{a,b=\pm}\sum_{\sigma,\sigma'}\int_0^Lc_{a,\sigma}^\dagger(ax) c_{b,\sigma'}^\dagger(bx) c_{a,\sigma'}(ax)
c_{b,\sigma}(bx)dx\\
+J\sum_{a=\pm}\sum_{\sigma,\sigma'}c_{a,\sigma}^\dagger(0) c_{a,\sigma'}(0){\bf \tau}_{\sigma,\sigma'}\cdot{\bf S}+V\sum_{a=\pm}c_{a,\sigma}^\dagger(0)
c_{a,\sigma}(0),\nonumber
\end{eqnarray}
where $a,b=\pm$ denote the right and left half-chains; $c_{a,\sigma}^\dagger$ ($c_{a,\sigma}$) are the creation (annihilation)
operators of the conduction electrons; $J<0$ (ferromagnetic) is the Kondo coupling constant; $V$ is the scalar potential induced
by the impurity; $c>0$ is the interaction constant of the conduction electrons; $L$ is the length of a half-chain which will
be put to infinity in the thermodynamic limit; ${\bf \tau}$ is the Pauli matrix and ${\bf S}$ is the moment of the impurity
with spin $S$. The interacting term in the Hamiltonian contains the $\delta$-function type repulsion of the electrons in the same
half-chain and the $\delta(x+y)$ type spin-exchange between electrons in the different half-chains. 
  For the electrons near 
the impurity, the latter is short ranged and has a clear physical meaning since both the Kondo coupling and the correlations in
the bulk assist such an exchange effect. In fact, as predicted
by the boundary conformal field theory\cite{20}, there is indeed a quasi-long-range ferromagnetic correlation between the
electrons on sites $n$ and $-n$ [Hamiltonian (1)],
\begin{eqnarray}
<{\sigma}_{n}\cdot{\sigma}_{-n}>\sim (2n)^{-\theta},
\end{eqnarray}
where the exponents $\theta$ varies from 1 to 4, which can be derived from the boundary conformal filed theory\cite{20}.
For the electrons far away from the impurity, the spin exchange interaction is somewhat
artificial but is irrelevant to the impurity behavior. Such an irrelevant term is conventionally introduced in the multi-channel
Kondo problem\cite{21} to ensure the model to be integrable. By reflecting the coordinates of the right half-chain, the
model is readily reduced to a two-channel Kondo model with a boundary impurity. The Hamiltonian (2) can be written in
the first quantization form as
\begin{eqnarray}
H=-\sum_{j=1}^N\frac{\partial^2}{\partial {x_j^2}}+c\sum_{i<j}^N(P_{ij}^c-P_{ij}^s)\delta(x_j-x_i)+\sum_{j=1}^N
(J\tau_j\cdot{\bf S}+V)\delta (x_j),
\end{eqnarray}
where $N$ is the number of electrons; $P_{ij}^c$ and $P_{ij}^s$ are the channel-channel and spin-spin exchange operators,
respectively; the coordinates of the electrons are constrained in the interval $-L\leq x_j\leq 0$. Without the impurity,
the Hamiltonian is just that of the two band model considered by Schlottmann\cite{22}.
\par
The structure of the present paper is the following. In the subsequent section, we derive the integrable condition of the
present model. The Bethe ansatz equation for the integrable case will be given with considerable detail. In section III, we
derive the thermodynamic Bethe ansatz equations based on the string hypothesis. The boundary bound states and the residual
entropy will be discussed in sect.IV. In Sect. V, we study the low-temperature thermodynamics of the impurity. Concluding 
remarks will be given in section VI.
\section{bethe ansatz}
It is well known that without the impurity, the Hamiltonian (3) is exactly solvable\cite{22}. By including the impurity, any
electron impinging on the impurity will be completely reflected and suffer a reflection matrix $R_j$. The waves are
therefore reflected at either end as
\begin{eqnarray}
e^{ik_jx}\to -e^{-ik_jx+2ik_jL}, {~~~}x\sim -L\nonumber\\
e^{ik_jx}\to R_j(k_j)e^{-ik_jx}, {~~~}x\sim 0.
\end{eqnarray}
Let us consider the two particle case. There are two ways from an initial state $(k_1,k_2,|)$ to a final state $(-k_1,-k_2,
|)$:
$$
I. (k_1,k_2,|)\to(k_2,k_1,|)\to(k_2,-k_1,|)\to(-k_1,k_2,|)\to(-k_1,-k_2,|),
$$
$$
II. (k_1,k_2,|)\to(k_1,-k_2,|)\to(-k_2,k_1,|)\to(-k_2,-k_1,|)\to(-k_1,-k_2,|),
$$
where the symbol $|$ denotes the open boundary. Since the physical process is unique, the following equation must hold
\begin{eqnarray}
S_{12}(k_1,k_2)R_1(k_1)S_{12}(k_1,-k_2)R_2(k_2)=R_2(k_2)S_{12}(k_1,-k_2)R_1(k_1)S_{12}(k_1,k_2).
\end{eqnarray}
Above $S_{12}$ is the scattering matrix between the two electrons. This is just the reflection equation
\cite{23}. For the multi-particle cases,
as long as the scattering matrix is factorizable or the two-body scattering matrix satisfies 
the Yang-Baxter relation\cite{24}
\begin{eqnarray}
S_{12}(k_1,k_2)S_{13}(k_1,k_3)S_{23}(k_2,k_3)=S_{23}(k_2,k_3)S_{13}(k_1,k_3)S_{12}(k_1,k_2),
\end{eqnarray}
(6) is the only constraint to the integrability of an open boundary system\cite{23}. 
 Below we derive the integrable condition of the present model.
\par
Since the reflection process only consists of one-electron effect, it is convenient to consider
the single-particle eigenstate. The Schr\"{o}dinger equation for this case reads
\begin{eqnarray}
-\frac{\partial^2\Psi(x)}{\partial x^2}+(J\tau \cdot{\bf S}+V)\delta (x)\Psi(x)=E\Psi(x).
\end{eqnarray}
We make the following ansatz for the wave function $\Psi(x)$
\begin{eqnarray}
\Psi(x)=[A_+e^{ikx}+A_-e^{-ikx}]\theta(-x)\theta(x+L),
\end{eqnarray}
where $\theta(x)$ is the step function which takes the value of unit for $x\geq 0$ and $0$ for $x<0$. For $-L<x<0$, we easily
obtain the eigenvalue $E$ takes the value $k^2$. Since the boundary contains an impurity, we use the open-string
 boundary condition, which has been used in a similar problem\cite{25}, to solve our model. That means  an irrelevant
local counter term $\sum_j[\delta'(x_j)-\delta'(x_j+L)]$ is included in the original Hamiltonian. For $x=-L$, we obtain 
that
\begin{eqnarray}
\frac{A_+}{A_-}=e^{2ikL}.
\end{eqnarray}
Similarly, for $x=0$, the $\delta(x)$ term must be canceled in (8). This gives
\begin{eqnarray}
R(k)\equiv\frac{A_-}{A_+}=\frac{2ik-(J\tau\cdot{\bf S}+V)}{2ik+(J\tau\cdot{\bf S}+V)}.
\end{eqnarray}
The self-consistent condition for $A_\pm$ constrains the value of $k$ by the following eigenvalue problem
\begin{eqnarray}
R(k)A_+=e^{-2ikL}A_+.
\end{eqnarray}
For arbitrary $N-$particle case, $R_j(k_j)$ must satisfy the reflection equation (6). It is known that the bulk Hamiltonian
is integrable\cite{22} and the the two-body 
scattering matrix of the the conduction electrons  takes
the form\cite{}
\begin{eqnarray}
S_{jl}(k_j-k_l)=\frac{k_j-k_l+icP_{jl}^s}{k_j-k_l+ic}\frac{k_j-k_l-icP_{jl}^c}{k_j-k_l-ic},
\end{eqnarray}
and obviously satisfy the Yang-Baxter equation (7). 
Substituting $S_{jl}$ of (13) and $R(k)$ of (11) into (6), we readily obtain the integrable condition of the present model. It reads
\begin{eqnarray}
J=-2c.
\end{eqnarray}
The reflection matrix (11) in the integrable case can be rewritten as
\begin{eqnarray}
R(k_j)=\frac{k_j+icp-ic{\bf \tau}_j\cdot{\bf S}}{k-icp+icS}\frac{k-ic(p+1)-ic{\bf \tau}_j\cdot{\bf S}}{k_j-ic(p+1)-icS},
\end{eqnarray}
where $p=V/2c$. Notice here $R(k_j)$ is an operator one which reveals the 
spin exchange process at the impurity site (boundary).
For an $N-$particle system, suppose the wave function initially has an amplitude $\psi_0$. When the $j-th$ particle moves 
across another particle, it gets an $S-$matrix $S_{jl}(k_j-k_l)$. At the right boundary, it is completely reflected back 
and suffer a factor
$R(k_j)$. Then it begins to move toward the left boundary. When reaches the left boundary, it will be kicked back and suffer
a factor $\exp(2ikL)$. Finally it arrives at the initial site and finishes a periodic motion. Therefore we have the following
equation
\begin{eqnarray}
S_{jj-1}^-\cdots S_{j1}^-S_{j1}^+\cdots S_{jj-1}^+S_{jj+1}^+\cdots S_{jN}^+R(k_j)e^{2ik_jL}S_{jN}^-\cdots S_{jj+1}^-
\psi_0=\psi_0,
\end{eqnarray}
or more neatly
\begin{eqnarray}
S_{jj-1}^-\cdots S_{j1}^-S_{j1}^+\cdots S_{jj-1}^+S_{jj+1}^+\cdots S_{jN}^+R(k_j)S_{jN}^-\cdots S_{jj+1}^-\psi_0=
e^{-2ik_jL}\psi_0,
\end{eqnarray}
where $S_{jl}^\pm=S_{jl}(k_j\pm k_l)$. Eq.(17) is just the reflection version of Yang's eigenvalue problem\cite{24}. Its
solution gives the Bethe ansatz equation. We note that the degrees of freedom of the spin sector and those of the channel
sector are completely separated. The scattering-matrices $S_{jl}$ can be expressed as
\begin{eqnarray}
S_{jl}^\pm=S_{jl}^{\pm,c}\otimes S_{jl}^{\pm,s},\nonumber\\
S_{jl}^{\pm,c}=\frac{k_j\pm k_l-icP_{jl}^c}{k_j\pm k_l-ic},\\
S_{jl}^{\pm,s}=\frac{k_j\pm k_l+icP_{jl}^s}{k_j\pm k_l+ic}.\nonumber
\end{eqnarray}
Therefore, the eigenvalue problem (17) is equivalent to
\begin{eqnarray}
S_{jj-1}^{-,s}\cdots S_{j1}^{-,s}S_{j1}^{+,s}\cdots S_{jj-1}^{+,s}S_{jj+1}^{+,s}\cdots S_{jN}^{+,s}
R(k_j)S_{jN}^{-,s}\cdots S_{jj+1}^{-,s}\psi_0^s=e_s(k_j)\psi_0^s,\\
S_{jj-1}^{-,c}\cdots S_{j1}^{-,c}S_{j1}^{+,c}\cdots S_{jj-1}^{+,c}S_{jj+1}^{+,c}\cdots S_{jN}^{+,c}
S_{jN}^{-,c}\cdots S_{jj+1}^{-,c}\psi_0^c=e_c(k_j)\psi_0^c,
\end{eqnarray}
with
\begin{eqnarray}
\psi_0=\psi_0^c\otimes\psi_0^s,{~~~~~~~~}e_c(k)e_s(k)=e^{-2ik_jL}.
\end{eqnarray}
The above eigenvalue problems are very similar to those of other integrable models\cite{25,26,27,23}. Here we introduce a
slightly different method, which is more transparent\cite{28}. Define ${\bar\psi}_0=(S_{jj-1}^{-,s}\cdots S_{j1}^{-,s})^{-1}
\psi_0^s$. $(19)$ can be rewritten as
\begin{eqnarray}
S_{j1}^{+,s}\cdots S_{jj-1}^{+,s}S_{jj+1}^{+,s}\cdots S_{jN}^{+,s}R(k_j)S_{jN}^{-,s}\cdots S_{jj+1}^{-,s}
S_{jj-1}^{-,s}\cdots S_{j1}^{-,s}{\bar\psi}_0
\equiv X_j{\bar\psi}_0=e_s(k_j){\bar \psi}_0.
\end{eqnarray}
For convenience, we introduce an auxiliary space $\tau$ and define
\begin{eqnarray}
U_\tau(k)=S_{\tau j}^{+,s}S_{\tau1}^{+,s}\cdots S_{\tau j-1}^{+,s}S_{\tau j+1}^{+,s}\cdots S_{\tau N}^{+,s}
R(k_\tau)S_{\tau N}^{-,s}
\cdots S_{\tau j+1}^{-,s}S_{\tau j-1}^{-,s}\cdots S_{\tau 1}^{-,s}S_{\tau j}^{-,s},
\end{eqnarray}
with $k_\tau=k$. Obviously, $S_{\tau j}^{-,s}(k_j)=P_{\tau j}^s$ and
\begin{eqnarray}
tr_\tau U_\tau(k_j)=\frac{2k_j+2ic}{2k_j+ic}X_j.
\end{eqnarray}
Since $S_{\tau l}^{\pm,s}$ satisfy the Yang-Baxter relation
\begin{eqnarray}
S_{\tau\tau'}^{-,s}(k-k')S_{\tau j}^{\pm,s}(k\pm k_j)S_{\tau' j}^{\pm,s}(k'\pm k_j)= S_{\tau' j}^{\pm,s}(k'\pm k_j)
S_{\tau j}^{\pm,s}(k\pm k_j)S_{\tau,\tau'}^{-,s}(k-k'),
\end{eqnarray}
from (6) we can easily show that $U_\tau (q)$ satisfies the reflection equation
\begin{eqnarray}
S_{\tau\tau'}^{-,s}(k-k')U_\tau (k)S_{\tau\tau'}^{+,s}(k+k')U_{\tau'}(k')=U_{\tau'}(k')S_{\tau\tau'}^{+,s}(k+k')
U_\tau (k)S_{\tau,\tau'}^{-,s}(k-k').
\end{eqnarray}
Therefore, the eigenvalue problem (22) is reduced to Sklyanin's eigenvalue problem\cite{23}. Following the standard method,
we obtain
\begin{eqnarray}
e_s^{-1}(k_j)=\frac{k_j-ic(S-p)}{k_j+ic(S-p)}\prod_{r=\pm}\prod_{\alpha=1}^M\frac{k_j-r\lambda_\alpha+i\frac c2}
{k_j-r\lambda_\alpha-i\frac c2},\\
\frac{\lambda_\alpha+ic(S+p+\frac12)}{\lambda_\alpha-ic(S+p+\frac12)}\frac{\lambda_\alpha+ic(S-p-\frac12)}
{\lambda_\alpha-ic(S-p-\frac12)}\prod_{j=1}^N \prod_{r=\pm}\frac{\lambda_\alpha-rk_j+i\frac c2}{\lambda_\alpha-rk_j-i\frac c2}
=\prod_{r=\pm}\prod_{\beta\neq\alpha}^M\frac{\lambda_\alpha-r\lambda_\beta+ic}{\lambda_\alpha-r\lambda_\beta-ic}.
\end{eqnarray}
Similarly, we have
\begin{eqnarray}
e_c^{-1}(k_j)=\prod_{r=\pm}\prod_{\gamma=1}^{\bar M}\frac{k_j-r\chi_\gamma-i\frac c2}{k_j-r\chi_\gamma+i\frac c2},\\
\prod_{j=1}^N\prod_{r=\pm}\frac{\chi_\gamma-rk_j+i\frac c2}{\chi_\gamma-rk_j+i\frac c2}=\prod_{r=\pm}\prod_{\delta\neq\gamma}^{\bar M}
\frac{\chi_\gamma-r\chi_\delta+ic}{\chi_\gamma-r\chi_\delta-ic}.
\end{eqnarray}
By using (21), we obtain
\begin{eqnarray}
e^{2ik_jL}=\frac{k_j-ic(S-p)}{k_j+ic(S-p)}\prod_{r=\pm}\{\prod_{\alpha=1}^M\frac{k_j-r\lambda_\alpha+i\frac c2}
{k_j-r\lambda_\alpha-i\frac c2}\prod_{\gamma=1}^{\bar M}\frac{k_j-r\chi_\gamma-i\frac c2}{k_j-r\chi_\gamma+i\frac c2}\}.
\end{eqnarray}
The energy spectrum of (4) is uniquely determined by the Bethe ansatz equations (BAE) (28), (30) and (31)
with the eigenvalue of (4) as
\begin{eqnarray}
E=\sum_{j=1}^Nk_j^2,
\end{eqnarray}
where $k_j$, $\lambda_\alpha$ and $\chi_\gamma$ are the rapidities of the charge, spin and channel (two half chains),
respectively; $M$ is the number of down spins and ${\bar M}$ is the number of the electrons
in the left half-chain. From the Bethe ansatz equation we can read off that the impurity spin does not behave as its real
value but as two ``images" or effective spins $S+p+1/2$ and $S-p-1/2$ alternatively and halfly. Since there is no constraint to the value
of $p$, we shall call the two effective spins as ghost spins. Notice that the mean value of the two ghost spins is still $S$.
\section{thermodynamic equations}
In this section, we derive the thermodynamic Bethe ansatz equations (TBAE) by following the standard method developed
by Yang and Yang\cite{29} and Takahashi\cite{30}. Due to the reflection symmetry of the model, $k_j$, $\lambda_\alpha$ and $\chi_\gamma$
are all positive parameters. Though the zero modes are possible solutions of the BAE, they corresponds to 
vanished wavefunction\cite{20}.
For convenience, we put $k_{-j}=-k_j$, $\lambda_{-\alpha}=-\lambda_\alpha$ and $\chi_{-\gamma}=-\chi_\gamma$. When $0\leq p
<S$ (weak impurity potential), an imaginary charge mode $k=i(S-p)c$ is a solution of the BAE in the thermodynamic limit
$L\to \infty$, while for $S\leq p$ (strong impurity potential), such a state is not allowed. The other $k$ modes then 
either take real values or form 
tightly bounded two-strings with the following fusion rule
\begin{eqnarray}
k_{\gamma}^{\pm}=\chi_\gamma\pm i\frac c2.
\end{eqnarray}
These modes correspond to the channel-singlet pairs. At low temperatures, the bound state mentioned above are much stable.
Since we are interested only in the low temperature behavior of the impurity, we shall omit the effect of breaking bound
state in deriving the free energy due to the finite energy gaps associated with such processes.
\par
(i). $S>p$ case. In this case, a boundary bound state with $k=ic(S-p)$ is stable. We remark that the electrons exactly on
the impurity site lose their channel character because this site may belong to either the left half-chain or the right
half-chain as we can see in the Hamiltonian (1). That means maximumly only two electrons with different spins may occupy
this site simultaneously and a pair state described by (33) to occupy the impurity site is forbidden. 
With the fusion rule (33),
The BAE can be reduced to
\begin{eqnarray}
e^{4i\chi_\gamma L}=
\prod_{r=\pm1}\frac{\chi_\gamma-ic(S-p+\frac12r)}{\chi_\gamma+ic(S-p+\frac12r)}
\prod_{\alpha=-M}^M\frac{\chi_\gamma-\lambda_\alpha+ic}{\chi_\gamma-\lambda_\alpha-ic}
\prod_{\delta=-{\bar N}}^{\bar N}\frac{\chi_\gamma-\chi_\delta-ic}{\chi_\gamma-\chi_\delta+ic},\\
\prod_{r=\pm1}\frac{\lambda_\alpha+ic(S+rp+\frac12)}{\lambda_\alpha-ic(S+rp+\frac12)}
\frac{\lambda_\alpha+\frac i2c}{\lambda_\alpha-\frac i2c}
\prod_{\gamma=-{\bar N}}^{\bar N}\frac{\lambda_\alpha-\chi_\gamma+ic}
{\lambda_\alpha-\chi_\gamma-ic}=\prod_{\beta=-M}^M\frac{\lambda_\alpha-\lambda_\beta+ic}
{\lambda_\alpha-\lambda_\beta-ic},
\end{eqnarray}
where ${\bar N}=(N-1)/2$ and the eigenenergy can be expressed as
\begin{eqnarray}
E=\sum_{\gamma=-{\bar N}}^{\bar N}{\chi_\gamma}^2 -Nc^2/2.
\end{eqnarray}
In the following text, we omit the second term in (36) since it only shifts the chemical potential. The solutions of the fused
BAE are described by a sequence of real number $\chi_\gamma$ and a variety of $\lambda-$strings
\begin{eqnarray}
\lambda_{\alpha,j}^n=\lambda_\alpha^n+\frac i2c(n+1-2j), {~~~~~~~}j=1,2,\cdots,n,
\end{eqnarray}
where $\lambda_\alpha^n$ are the position of the $\alpha$-th $n-$string in the real axis. The distribution of $\chi_\gamma$
and $\lambda_\alpha^n$ in the thermodynamic limit can be described by the densities of particles $\rho(\chi)$, $\sigma_n(\lambda)$
and the densities of holes $\rho_h(\chi)$ and $\sigma_n^h(\lambda)$, respectively\cite{30}. From the BAE we can easily derive the 
following equations
\begin{eqnarray}
\rho(\lambda)+\rho_h(\lambda)=\frac 1\pi+\frac 1{2L}\phi_c(\lambda)+[2]\rho(\lambda)-\sum_{n=1}^\infty{ B}_{n2}\sigma_n(\lambda),\\
\sigma_n^h(\lambda)=\frac1{2L}[\phi_s^n(\lambda)+\phi_b^n(\lambda)]+{ B}_{n2}\rho(\lambda)-\sum_{m=1}^\infty { A}_{nm}\sigma_m(\lambda),
\end{eqnarray}
with the condition
\begin{eqnarray}
\int\rho(\lambda)d\lambda=\frac N{2L}, {~~~~~~~}\sum_n n\int\sigma_n(\lambda)d\lambda=\frac{2M+1}{2L},
\end{eqnarray}
where
\begin{eqnarray}
{ B}_{mn}=\sum_{l=1}^{min(m,n)}[n+m+1-2l],\\
{ A}_{m,n}=[|m-n|]+2[|m-n|+2]+\cdots+2[m+n-2]+[m+n],
\end{eqnarray}
and $[n]$ is an integral operator with the kernel
\begin{eqnarray}
a_n(\lambda)=\frac1\pi\frac{\frac{nc}2}{\lambda^2+(\frac{nc}2)^2},
\end{eqnarray}
$\phi_c$, $\phi_s^n$ and $\phi_b^n$ are given by
\begin{eqnarray}
\phi_c(\lambda)=\sum_{r=\pm1}a_{2S-2p+r}(\lambda),\\
\phi_s^n(\lambda)=\sum_{r=\pm}\sum_{l=1}^na_{|n+2S+2+2rp-2l|}(\lambda)sgn(n+2S+2+2rp-2l),\\
\phi_b^n(\lambda)=a_n(\lambda).
\end{eqnarray}
The free energy of the system in a magnetic field $H$ can be expressed as\cite{29,30}
\begin{eqnarray}
F/L=\int(\lambda^2-\mu-H)\rho(\lambda)d\lambda+\sum_nnH\int\sigma_n(\lambda)d\lambda-(S+\frac12)H\nonumber\\
-T\int[(\rho+\rho_h)\ln(\rho+\rho_h)-\rho\ln\rho-\rho_h\ln\rho_h]d\lambda\\-T\sum_n\int[(\sigma_n+\sigma_n^h)\ln(\sigma_n+
\sigma_n^h)-\sigma_n\ln\sigma_n-\sigma_n^h\ln\sigma_n^h]d\lambda,\nonumber
\end{eqnarray}
where $\mu $ is the chemical potential. At equilibrium state, $\delta F=0$. That means
\begin{eqnarray}
\frac{\delta F}{\delta\rho(\lambda)}=\frac{\delta F}{\delta \sigma_n(\lambda)}=0.
\end{eqnarray}
Notice that $\delta\rho_h$ and $\delta\sigma_n^h$ are not independent of $\delta\sigma_n$ and $\delta \rho$. From (38) and (39)
we know that
\begin{eqnarray}
\delta\rho_h=-\delta\rho+[2]\delta\rho-\sum_n{ B}_{n2}\delta\sigma_n,\\
\delta\sigma_n^h={ B}_{n2}\delta\rho-\sum_m{ A}_{mn}\delta\sigma_n.
\end{eqnarray}
Therefore we have the following equations
\begin{eqnarray}
\ln\eta=\frac{\lambda^2-\mu-H}T-[2]\ln(1+\eta^{-1})-\sum_n{ B}_{n2}\ln(1+\zeta_n^{-1}),\\
\ln(1+\zeta_n)=\frac{nH}T+{ B}_{n2}\ln(1+\eta^{-1})+\sum_m{ A}_{nm}\ln(1+\zeta_m^{-1}),
\end{eqnarray}
where
\begin{eqnarray}
\eta(\lambda)\equiv\frac{\rho_h(\lambda)}{\rho(\lambda)},{~~~~~~}\zeta_n(\lambda)\equiv\frac{\sigma_n^h(\lambda)}{\sigma_n(\lambda)}.
\end{eqnarray}
Define the integral operator ${ G}$ with kernel $G_0(\lambda)=1/2c\cosh(\pi\lambda/c)$. It can be shown that the following identities are
valid
\begin{eqnarray}
B_{1n}-GB_{2n}=\delta_{n1}G,{~~~~~~}B_{m,n}-G(B_{n-1,m}+B_{n+1,m})=\delta_{nm}G,{~~}n>1,\\
A_{1m}-GA_{2m}=\delta_{1m},{~~~~~~}A_{nm}-G(A_{n-1,m}+A_{n+1,m})=\delta_{nm},{~~}n>1,\\
GA_{mn}=B_{mn}.
\end{eqnarray}
With the above identities, (51) and (52) can be further simplified. Let $n=2$ in (52) and act $G$ on it. We get
\begin{eqnarray}
\sum_nB_{n2}\ln(1+\zeta_n^{-1})=G\ln(1+\zeta_2)-[2]\ln(1+\eta^{-1})-\frac HT.
\end{eqnarray}
Substituting (57) into (51) we obtain
\begin{eqnarray}
\ln\eta=\frac{\lambda^2-\mu}T-G\ln(1+\zeta_2).
\end{eqnarray}
Similarly, with (54) and (55), (52) can be reduced to
\begin{eqnarray}
\ln\zeta_n=-G\ln(1+\eta^{-1})\delta_{n2}+G[\ln(1+\zeta_{n+1})+\ln(1+\zeta_{n-1})].
\end{eqnarray}
Notice we have put $\zeta_0\equiv 0$. (51), (52) and (58), (59) are just the TBAE. In the following text, we shall use them alternatively.
Substituting (51) and (52) into (47) we obtain the free energy as
\begin{eqnarray}
F/L=f+\frac1LF_i+\frac1LF_b,\\
f=-\frac T\pi\int \ln[1+\eta^{-1}(\lambda)]d\lambda,\\
F_i=-\frac12T\sum_n\int\phi_n^s(\lambda)\ln[1+\zeta_n^{-1}(\lambda)]d\lambda-\frac12T\int\phi_c(\lambda)
\ln[(1+\eta^{-1}(\lambda)]d\lambda-H(S+\frac12),\\
F_b=-\frac12T\sum_n\int a_n(\lambda)\ln(1+\zeta_n^{-1})d\lambda=-T\int\frac{\ln[1+\zeta_1(\lambda)]}{4c\cosh\frac{\pi\lambda}c}d\lambda.
\end{eqnarray}
\par
(ii)$S\leq p$ case. In this case, the boundary mode $k=i(S-p)c$ is not a solution of the BAE. The fused BAE read
\begin{eqnarray}
e^{4i\chi_\gamma L}=\prod_{r=\pm1}\frac{\chi_\gamma-ic(S-p-\frac12r)}{\chi_\gamma+ic(S-p-\frac12r)}
\prod_{\alpha=-M}^M\frac{\chi_\gamma-\lambda_\alpha+ic}{\chi_\gamma-\lambda_\alpha-ic}
\prod_{\delta=-{\bar N}}^{\bar N}\frac{\chi_\gamma-\chi_\delta-ic}{\chi_\gamma-\chi_\delta+ic},\\
\prod_{r=\pm1}\frac{\lambda_\alpha+ic(S+rp+\frac12r)}{\lambda_\alpha-ic(S+rp+\frac12r)}
\frac{\lambda_\alpha+\frac i2c}{\lambda_\alpha-\frac i2c}
\prod_{\gamma=-{\bar N}}^{\bar N}\frac{\lambda_\alpha-\chi_\gamma+ic}
{\lambda_\alpha-\chi_\gamma-ic}=\prod_{\beta=-M}^M\frac{\lambda_\alpha-\lambda_\beta+ic}
{\lambda_\alpha-\lambda_\beta-ic},
\end{eqnarray}
where ${\bar N}=N/2$. The TBAE take the same forms to those of case (i). The only difference is the free energy of the impurity,
which is given by a new class of $\phi_s^n$
\begin{eqnarray}
\phi_s^n(\lambda)=\sum_{r=\pm}\sum_{l=1}^na_{|n+2S+1+r(2p+1)-2l|}(\lambda)sgn(n+2S+1+r(2p+1)-2l).
\end{eqnarray}
\section{ground state properties}
To study the finite temperature properties of the system, we check first the behavior of the driving term in (59). Since we are 
interested in the low-temperature properties of the impurity, only the excitations near the Fermi surface are important. As
we discussed above, $\eta$ is relevant to the charge excitations and $\zeta_n$ are responsible for the spin excitations. Therefore,
when $T\to 0$ and $H\to 0$, only $\eta(\lambda\sim\pm\Lambda)$ and $\zeta_n(\lambda\sim\pm\infty)$ are important in the driving
term of (59). Introduce the notation
\begin{eqnarray}
\eta(\lambda,T)=e^{\frac{\epsilon_c(\lambda,T)}T}.
\end{eqnarray}
$\epsilon_c(\lambda,0)$ is nothing
but the dressed energy\cite{31} of the charge sector. At zero temperature, the Fermi surface of the charge sector is given by
$\lambda=\pm\Lambda$ with $\epsilon_c(\pm\Lambda,0)=0$. For $|\lambda|<\Lambda$, $\epsilon_c(\lambda,0)<0$ and
for $|\lambda|>\Lambda$, $\epsilon_c(\lambda,0)>0$. Based on the above argument, we have
\begin{eqnarray}
\ln(1+\eta^{-1})\approx -\frac{\epsilon_c(\lambda)}T, {~~~~}|\lambda|<\Lambda,\\
\ln(1+\eta^{-1})\approx e^{-\frac{\epsilon_c(\lambda)}T}, {~~~~} |\lambda|>\Lambda.
\end{eqnarray}
Therefore, the driving term in (59) can be expressed in its leading order as
\begin{eqnarray}
-G\ln(1+\eta^{-1})\approx -\frac D{2T\cosh\frac\pi c\lambda},{~~~~}|\lambda|\to\infty
\end{eqnarray}
where $D$ is the effective band width of the spinons. Thus at very low temperatures, (59) can be rewritten as
\begin{eqnarray}
\ln\zeta_n=-\frac D{2T\cosh\frac\pi c\lambda}\delta_{n,2}+G[\ln(1+\zeta_{n+1})+\ln(1+\zeta_{n-1})].
\end{eqnarray}
Such an equation is exactly the same to that of the spin-1 chain with an impurity\cite{19} and much similar to that of the 
conventional two-channel Kondo problem\cite{21} but with two Fermi points. As $T\to 0$, the driving term in (71) is divergent. That means
$\zeta_2\to 0$. In this case, all $\zeta_n$ are variable-independent and tend to a series of constants $\zeta_n^+$ and
$G\to 1/2$. Therefore, (71) is reduced to the following algebraic equations
\begin{eqnarray}
{\zeta_n^+}^2=(1+\zeta_{n+1}^+)(1+\zeta_{n-1}^+),
\end{eqnarray}
with the boundary conditions
\begin{eqnarray}
\zeta_0^+=\zeta_2^+=0,{~~~~~~~~}\lim_{n\to\infty}\frac{\ln\zeta_n^+}n=\frac HT.
\end{eqnarray}
As discussed
in the earlier publications\cite{30,21}, we have the following solution
\begin{eqnarray}
\zeta_n^+=\frac{\sinh^2[(n-1)x_0]}{\sinh^2x_0}-1, {~~~~}for {~~}n\geq 2,\\
\zeta_1^+=1,
\end{eqnarray}
where $x_0=H/2T$. Below we discuss the ground state properties of the impurity and the open boundary for different $p$ values.
First we consider the open boundary. Substituting (75) into (63) we find that the free energy of the open boundary takes the following
form
\begin{eqnarray}
F_b\to-\frac14T\ln2.
\end{eqnarray}
That means the residual entropy of the open boundary is
\begin{eqnarray}
S_b=\frac14\ln2.
\end{eqnarray}
Such a result is nothing but one half of that of a spin-1/2 in a two-channel electron system\cite{21}. We remark that the spin-1/2 degree
of freedom has been observed experimentally in a quasi-one-dimensional spin-1 system\cite{32}. In an open boundary system, the self
avoiding of scattering of a particle with its reflection wave leaves a spin-1/2 hole in the bulk. This can be readily read off
from the BAE. Such an effect induces the boundary degree of freedom and seems to be universal in the one-dimensional
open-boundary systems and does not depend on the detail of the impurity.
\par
Now we turn to the impurity. (i)$p<S$ case. In this case, when $2p=integer$, the local composite (local spin plus the bounded
electron) behaves as two ghost spins $S+p+1/2$ and $S-p+1/2$.
\begin{eqnarray}
\phi_n^s=a_{n,2S+2p+1}(\lambda)+a_{n,2S-2p+1}(\lambda),
\end{eqnarray}
where $a_{n,m}$ is the kernel of $B_{nm}$. With (56) and (52), we can rewrite (62) as
\begin{eqnarray}
F_i=F_i^s+F_i^c,\\
F_i^s=-T\int\frac1{4c\cosh\frac{\pi\lambda}c}[\ln(1+\zeta_{2S+2p+1})+\ln(1+\zeta_{2S-2p+1})]d\lambda,\\
F_i^c=-\frac12T\int\phi_c(\lambda)\ln(1+\eta^{-1})d\lambda+\sum_{r=\pm}GB_{2S+2rp+1,2}\ln[1+\eta^{-1}(0)].
\end{eqnarray}
With (74) we obtain the spin part of $F_i$ as
\begin{eqnarray}
F_i^s\to-\frac14T\ln[(1+\zeta_{2S+2p+1}^+)(1+\zeta_{2S-2p+1}^+)].
\end{eqnarray}
The residual entropy is
\begin{eqnarray}
S_{res}=-\lim_{T\to0}\lim_{H\to0}\frac{\partial F_i^s}{\partial T}=\ln[2\sqrt{S^2-p^2}]. 
\end{eqnarray}
In another hand, the residual magnetization can be derived as
\begin{eqnarray}
M_{res}=-\lim_{H\to0}\lim_{T\to0}\frac{\partial F_i^s}{\partial H}=S-\frac12.
\end{eqnarray}
When $S\geq 1$, the leading term of the impurity susceptibility is a Curie type since the moment of the local composite
can not be completely screened in this case. It can be calculated as
\begin{eqnarray}
\chi_i=-\lim_{H\to0}\frac{\partial^2F_i^s}{\partial H^2}=\frac{S^2+p^2-\frac14}{3T}+o(T^0).
\end{eqnarray}
Obviously, the effective moment of the impurity spin is enlarged by the impurity potential
\begin{eqnarray}
\mu_{eff}=\sqrt{S^2+p^2}.
\end{eqnarray}
Notice above we have put the Bohr magneton $\mu_B$ as our unit.
For $2p\neq integer$, we put $2{\bar p}=integer{~} part{~} of{~} 2p$. Since when $T\to 0$, $\zeta_n$ are almost variable-independent, 
$a_n(\lambda)$ in the integral can be replaced by $\delta(\lambda)$. In this case, we have
\begin{eqnarray}
\phi_n^s(\lambda)\to a_{n,2S+2{\bar p}+1}(\lambda)+a_{n,2S-2{\bar p}+1}(\lambda)+\delta(\lambda)\sum_{l=1}^{2{\bar p}}
\delta_{n,2S-2{\bar p}+2l}.
\end{eqnarray}
Therefore,
\begin{eqnarray}
S_{res}=\ln[2\sqrt{S^2-{\bar p}^2}]+\sum_{l=1}^{2{\bar p}}\ln\frac{2S-2{\bar p}+2l-1}{\sqrt{(2S-2{\bar p}+2l-1)^2-1}},\\
M_{res}=S-\frac12.
\end{eqnarray}
(ii)$p\geq S$ case. In this case, the boundary charge mode is not allowed due to the large impurity potential $V$. For
$2p=integer$, the impurity spin behaves as two ghost spins $S+p+1/2$ and $S-p-1/2$.
\begin{eqnarray}
\phi_n^s=a_{n,2S+2p+1}-a_{n,2p+1-2S}.
\end{eqnarray}
With the same procedure discussed above, we have
\begin{eqnarray}
S_{res}=\ln\sqrt{\frac{p+S}{p-S}}, {~~~~}p>S,\\
M_{res}=S, {~~~~~}p>S.
\end{eqnarray}
Notice the above relation is not valid for $p=S$. In this case
\begin{eqnarray}
S_{res}=\frac12\ln[2\sqrt2S],{~~~~~~}M_{res}=S-\frac14.
\end{eqnarray}
For $2p\neq integer$,
\begin{eqnarray}
\phi_n^s\to a_{n,2S+2{\bar p}+1}-a_{n,2{\bar p}+1-2S}+\delta(\lambda)\sum_{l=1}^{2S}\delta_{n,2{\bar p}-2S+2l},
\end{eqnarray}
and
\begin{eqnarray}
S_{res}=\ln\sqrt{\frac{{\bar p}+S}{{\bar p}-S}}+\sum_{l=1}^{2S}\ln\frac{2{\bar p}-2S+2l-1}
{\sqrt{(2{\bar p}-2S+2l-1)^2-1}},
{~~~}{\bar p}>S\\
S_{res}=\frac12\ln S+\sum_{l=2}^{2S}\ln\frac{2l-1}{\sqrt{4l(l-1)}},{~~~~~}{\bar p}=S.
\end{eqnarray}
The residual magnetization takes the same form of (89) and (93).
\par
From the above discussion, we obtain that the impurity potential has a significant effect to the spin dynamics of the impurity.
Generally, it splits the impurity spin into two ghost spins. For a weak impurity potential $V$ ($p<S$), the Furusaki-Nagaosa
conjecture is perfectly realized (Notice that $M_{res}=0$ for $S=1/2$, which indicates a ferromagnetic Kondo screening). While
for $p\geq S$, a non-zero residual magnetization always exists even for $S=1/2$. That means the Furusaki-Nagaosa's singlet 
is unstable in this case. In fact, there is a competition between $J$ and $V$, which governs the stability of the ferromagnetic
Kondo screening in this system. In addition, the residual entropy strongly depends on the impurity potential. Such an effect
characters the local glass behavior near the impurity. Notice that the residual entropy is disconnected at $2p=integer$. That
means quantum phase transition occurs when $2p$ across an integer. All the novel phenomena mentioned above have never been
obtained in the conventional Kondo problems.
\section{low-temperature properties}
 To get the leading order of the thermodynamics, we expanse (71) in the limit $T\to 0$. In this case, the excitations near
 $|\lambda|\to\infty$ dominates and the driving term can be approximately expressed as $D/T\exp(-\frac\pi c|\lambda|)$. For convenience, we introduce the new variables
$z_\pm=\pm\lambda\pi/c+\ln(D/T)$. With the method developed by Andrei and coworkers\cite{21}, we know that $\zeta_n$ allow the 
following asymptotic expansions
\begin{eqnarray}
\zeta_1(z_\pm)=1+(\alpha_1+\beta_1x_0^2)z_\pm e^{-z_\pm}+\cdots,\\
\zeta_n(z_\pm)=\zeta_n^++(\alpha_n+\beta_nx_0^2)e^{-z_\pm}+\cdots, {~~~~~}n\geq2.
\end{eqnarray}
Here $\alpha_n$ and $\beta_n$ are some constants. With the above expressions, the temperature dependences of some thermodynamic
quantities can be easily derived. First we consider the open boundary effect. Substituting (97) into (63), we readily obtain
that $\ln T$ terms appear in the free energy, which give the specific heat and the susceptibility of the open boundary as
\begin{eqnarray}
C_b\sim-T\ln T, {~~~~~~}\chi_b\sim -\ln T.
\end{eqnarray}
The above result indicates a typical overscreened two-channel Kondo behavior. Below we discuss the low-temperature behavior
of the impurity for different $p$ values.
\par
(i)$0\leq p<1/2$ case. This case may provide us clear information about how the impurity potential behaves at low temperatures.
In this case, $\phi_n^s$ can be written as
\begin{eqnarray}
\phi_n^s=\sum_{r=\pm}a_{n,2S+1}(\lambda+ircp).
\end{eqnarray}
Therefore, the spin part of the free energy of the impurity takes the form
\begin{eqnarray}
F_i^s=-\frac12T\sum_{r=\pm}\int\frac{\ln[1+\eta_{2S+1}(\lambda)]}{2c\cosh\frac\pi c(\lambda+ircp)}d\lambda\nonumber\\
=-\frac T{c}\int\frac{\cosh\frac\pi c\lambda\cos(p\pi)}{\cosh\frac{2\pi}c\lambda+\cos(2p\pi)}\ln[1+\eta_{2S+1}(\lambda)]d\lambda,
\end{eqnarray}
Substituting (98) into the above equation, we obtain that
\begin{eqnarray}
F_i^s=-\frac12T\ln(1+\zeta_{2S+1}^+)-AT^2-BH^2+o(T^2,H^2),
\end{eqnarray}
where $A,{~}B$ are two positive constants. For $S=1/2$, we obtain the specific heat $C_i^s$ and the susceptibility $\chi_i$
as
\begin{eqnarray}
C_i^s=2AT,{~~~~~~~}\chi_i=2B,
\end{eqnarray}
which indicate a typical local Fermi liquid behavior. Notice no overscreened two-channel Kondo effect exists in this case because
of the formation of the local composite. However, the Furusaki-Nagaosa's conjecture is perfectly realized. When $S\geq 1$, a
Schottky term and a Curie term appear in the specific heat and susceptibility, respectively. That means the local composite
is underscreened. Notice that (101) is correct only for $p<1/2$. When $p\to 1/2$, (101) tends to zero, which indicates a quantum phase
transition at this point. In fact, when $p=1/2$, the local composite behaves as two effective spins $S+1$ and $S$ ($S\geq 1$).
Such a spin-splitting effect drives the system  toward a different fixed point.
\par
(ii)$0<p<S, {~}2p=integer$ case. In this case, the local composite is split into two ghost spins $S+p+1/2$ and $S-p+1/2$.
\begin{eqnarray}
F_i^s=-\frac12T\sum_{r=\pm}\int G_0(\lambda)\ln[1+\zeta_{2S+2rp+1}(\lambda)]d\lambda.
\end{eqnarray}
Substituting (98) into the above equation, we can easily derive that
\begin{eqnarray}
F_i^s=-\frac14T\ln[(1+\zeta_{2S+2p+1}^+)(1+\zeta_{2S-2p+1}^+)]-A'T^2-B'H^2+o(T^2,H^2),
\end{eqnarray}
where $A'$ and $B'$ are two positive constants. The low-temperature behavior is much similar to that of case (i).
\par
(iii)$p=S$ case. In this case, no boundary bound state of charge exists. The local spin is split by the impurity potential
into an effective spin $2S+1/2$ and a spin-1/2 hole. The contribution of the spin hole is exactly canceled by the boundary spin
as we can see from the BAE (65). Therefore, the impurity and the boundary effect behave as effectively a spin $2S+1/2$. However,
its contribution to the thermodynamic quantities is only one half of that of a spin-$2S+1/2$ due to the reflection symmetry.
\par
(iv)$p<S,{~}2p\neq integer$ case. In this case,
\begin{eqnarray}
\phi_n^s(\lambda)=[\alpha]a_{n,2S+2{\bar p}+1}(\lambda)+[\alpha]^{-1}a_{n,2S-2{\bar p}+1}(\lambda)+a_\alpha(\lambda)\sum_{l=1}^{2{\bar p}}\delta_{n,2S-
2{\bar p}+2l},
\end{eqnarray}
where $\alpha=2(p-{\bar p})$. With the above relation we obtain
\begin{eqnarray}
F_i^s=-\frac12T\sum_{r=\pm}\int G_r(\lambda)\ln[1+\zeta_{2S+2r{\bar p}+1}(\lambda)]d\lambda\nonumber\\
-\frac12T\sum_{l=1}^{2{\bar p}}\int a_\alpha(\lambda)\ln[1+\zeta_{2S-2{\bar p}+2l}^{-1}(\lambda)]d\lambda,
\end{eqnarray}
where
\begin{eqnarray}
G_\pm(\lambda)=\int\frac{e^{\mp(p-{\bar p})|\omega|c}}{4\pi\cosh\frac{\omega c}2}e^{-i\lambda\omega}d\omega.
\end{eqnarray}
Notice $G_\pm(\lambda)$ is convergent in the real axis since $\alpha<1$. Since (107) is only related to $\zeta_n,{~}n\geq 2$, we
have a similar result of (105). That means the system falls the same universal class of $p={\bar p}$.
\par
(v)${\bar p}=S, {~}p\neq S$ case. No boundary bound state exists via the strong impurity potential.
\begin{eqnarray}
\phi_n^s(\lambda)=[\alpha]a_{n,4S+1}-[\alpha]^{-1}a_n+a_\alpha\sum_{l=1}^{2S}\delta_{n,2l}.
\end{eqnarray}
With the similar procedure we have
\begin{eqnarray}
F_i^s=-\frac12T\int G_+\ln(1+\zeta_{4S+1})d\lambda+\frac12T\int G_-\ln(1+\zeta_1)d\lambda\nonumber\\
-\frac12T\sum_{l=1}^{2S}\int a_\alpha(\lambda)\ln(1+\zeta_{2l}^{-1})d\lambda.
\end{eqnarray}
Very interestingly, $\zeta_1$ appears in the above expression. From (97) we know that it gives a $T^2\ln T$ term in $F_i^s$. 
This induce the negative specific heat and susceptibility
\begin{eqnarray}
C_i^s\sim T\ln T, {~~~~~~}\chi_i\sim \ln T.
\end{eqnarray}
However, we know that the contributions of the open boundary always take the same form of (111)
 but with negative sign and larger amplitudes. Therefore, the total specific heat 
and the susceptibility are still positive, i.e., well defined.
\par
(vi)${\bar p}>S,{~} 2p\neq integer$ case. In this case, the local spin is split into a ghost spin $S+{\bar p}+1/2$ and a ghost
spin hole ${\bar p}-S+1/2$.
\begin{eqnarray}
\phi_n^s(\lambda)=[\alpha]a_{n,2S+2{\bar p}+1}(\lambda)-[\alpha]^{-1}a_{n,2{\bar p}-2S+1}(\lambda)+a_\alpha(\lambda)\sum_{l=1}^{2S}
\delta_{n,2{\bar p}-2S+2l}.
\end{eqnarray}
The free energy takes the form
\begin{eqnarray}
F_i^s=-\frac12T\sum_{r=\pm}\int rG_r\ln(1+\zeta_{2{\bar p}+2rS+1})d\lambda-\frac12T\sum_{l=1}^{2S}\int a_\alpha\ln(1+\zeta_{2{\bar p}-2S+2l}^{-1})
d\lambda.
\end{eqnarray}
Since only $\zeta_n$ with $n\geq 2$ are related to the above expression. The thermodynamic behavior of this case is much similar
to that of case (iv).

\section{concluding remarks}
In conclusion, we propose an integrable model for the ferromagnetic Kondo problem in an interacting one-dimensional electron system.
 The impurity potential, originally
 a charge effect, enters the spin dynamics effectively at low energy scales. It is found that there is a competition between
 the Kondo coupling $J$ and the impurity potential $V$. When the ferromagnetic Kondo coupling dominates over the impurity potential
($V<|SJ|$), the Furusaki-Nagaosa many-body singlet can be perfectly realized. For example, when $S=1/2$, the residual magnetization
of the impurity is zero. However, when the impurity potential
dominates over the Kondo coupling ($V\geq |SJ|$), the fixed point predicted by Furusaki and Nagaosa is unstable
and the system must flow to a weak coupling fixed point. When $S\leq p<S+1/2$, the local moment is partially screened
($M_{res}=S-1/4$) and when $p\geq S+1/2$, the impurity moment can not be screened any more by the conduction electrons
($M_{res}=S$). Such a phenomenon is very different from that of the conventional Kondo problem
 in a Fermi liquid, where the impurity
potential only renormalizes the Kondo coupling constant but does not affect the fixed point of the system. Generally, the 
impurity potential splits the impurity spin into two ghost spins, which are responsible to the thermodynamic behavior
of the impurity. The effective moment of the impurity measured from the low temperature susceptibility is considerably enlarged
by the scalar potential. It is found that the residual entropy depends not only on the residual magnetization of the 
impurity, but also on the value of the impurity potential, which reveals the local spin glass behavior near the impurity.
At $2p=integers$, the ghost spins change their values and the residual entropy has a jump. That means  quantum phase transitions
occur at these points. Due to the formation of the local composite in the case of weak impurity potential, no overscreened
two channel Kondo behavior exists for the impurity. However, the open boundary behaves always like a spin 1/2 (halfly) and 
therefore shows overscreened two-channel Kondo behavior at low temperatures. In this paper, we consider only the $V>0$ case.
For a negative impurity potential, say $p=-S$, the local spin is split into two ghost spins $2S+1/2$ and $1/2$. No matter how
large the impurity spin is, the small ghost spin may induce the overscreened two-channel Kondo behavior. In this case, the
leading term in the susceptibility is Curie type while that in the low-temperature specific heat is $\sim -T\ln T$. 
 Based on the open boundary condition at the impurity site, we do not obtain
the non-Fermi liquid behavior predicted in [8]. It has been shown by the boundary conformal field theory\cite{20} that such
 a non-Fermi
liquid behavior is induced by the tunneling of the conduction electrons through the impurity. However, the tunneling introduces
a hybridization between the electrons in the two different half-chains and must induce the channel anisotropy and splitting. Generally, the 
tunneling is much weaker than the Kondo coupling and the impurity potential in a repulsive interaction system and is thus not
very harmful to the two-channel Kondo effect\cite{33,34,17}. In addition, it is not available to recover Lee and Toner's
conjecture about the power-law dependence of the Kondo temperature on the Kondo coupling constant $J$ from the present
model since the ratio $J/c$ is fixed in our model.  We remark that the antiferromagnetic Kondo problem in  multi-channel
correlated host was considered by a few authors\cite{35,15} recently. For a magnetic impurity coupled antiferromagnetically with
a Luttinger liquid, Egger and Komnik found the overscreened two-channel Kondo effect at low temperatures\cite{15}. However,
Zvyagin and Schlottmann showed in a recent paper\cite{35} that the multi-channel Kondo behavior may be smeared by
the correlation in the host. It is instructive to apply the RG analysis and the conformal field
theory on this problem to give a full picture of scalar potential effect to the Kondo problem (both ferromagnetic and
antiferromagnetic) in  Luttinger liquids as well as in higher-dimensional correlated systems.
\par
One of the authors (Y.W.) acknowledges the financial aids of Alexander von Humboldt-Stiftung and China National Foundation
of Natural Science.

\end{document}